\documentclass[epjCONF]{svjour}
\usepackage{graphics}
\usepackage[varg]{txfonts} % Times fonts
\usepackage[latin1]{inputenc}
\session-title{Conference Title, to be filled}
\begin{document}
%contributed talks: 6 pags.

\title{Pulsations driven by the $\epsilon$-mechanism in post-merger
  remnants: first results} \author{Marcelo M. Miller
  Bertolami\inst{1,2,3}\fnmsep\thanks{\email{mmiller@fcaglp.unlp.edu.ar}}
  \and Alejandro H. C\'orsico \inst{1,2} \and Xianfei Zhang \inst{4}
  \and Leandro G. Althaus\inst{1,2} \and C. Simon Jeffery \inst{4,5}}
\institute{Facultad de Ciencias Astron\'omicas y Geof\'isicas,
  Universidad Nacional de La Plata, Paseo del Bosque s/n, 1900 La
  Plata, Argentina.\and CONICET, Paseo del Bosque s/n, 1900 La
  Plata, Argentina.\and Max-Planck-Institut f\"ur
  Astrophysik, Karl-Schwarzschild-Str. 1, 85748, Garching,
  Germany.\and Armagh Observatory, College Hill, Armagh BT61 9DG,
  UK. \and School of Physics, Trinity College Dublin, Dublin 2,
  Ireland}
\abstract{Helium-rich subdwarfs are a rare subclass of hot subdwarf
  stars which constitute a small and inhomogeneous group showing
  varying degrees of helium enrichment. Only one star, LS IV
  $^\circ$14 116 has been found to show multiperiodic luminosity
  variations. The variability of LS IV $^\circ$14 116 has been
  explained as the consequence of nonradial g-mode oscillations, whose
  excitation is difficult to understand within the frame of the
  standard $\kappa$-mechanism driving pulsations in sdBV stars. In a
  recent study, we have proposed that the pulsations of LS IV
  $^\circ$14 116 might be driven through the $\epsilon$-mechanism
  acting in unstable He-burning zones in the interior of the star,
  that appear before the quiescent He-burning phase. One of the few
  accepted scenarios for the formation of He-rich subdwarfs is the
  merger of two He-core white dwarfs. As part of this project, we
  present a study of the $\epsilon$-mechanism in post-merger remnants,
  and discuss the results in the light of the pulsations exhibited by
  LS IV $^\circ$14 116.} %end of abstract
\maketitle
\section{Introduction}
\label{intro}

About 5\% of all hot subdwarf stars (sdB, sdO) show helium
(He)-enriched surface abundances (He-sdB, He-sdO). While most normal
hydrogen (H)-rich subdwarfs are supposed to be low mass core
He-burning stars with atmospheres dominated by H due to the action of
gravitational settling, the evolutionary status of the He-rich
subclass is less clear. He-rich subdwarfs have been suggested to be
the result of either double He white dwarf (HeWD) mergers
\cite{2000MNRAS.313..671S} or late helium core flashes
(a.k.a. hot-flashers\cite{2001ApJ...562..368B}). In fact, it is by no
means clear whether intermediate and extreme He-rich sdB stars form a
single class of objects or not \cite{2012Jeffery}. Regardless of the
particular scenario proposed for their formation, it is accepted that
some of these stars are still contracting towards the He-core burning
phase (EHB), as otherwise gravitational settling of the remaining H
would have turned the star into a H-rich sdB star.

Two families of pulsators have been discovered within the H-rich sdB
stars: the rapid pulsators (sdBVr; \cite{2010IBVS.5927....1K})
discovered by \cite{1997MNRAS.285..640K}, and the slow pulsators
(sdBVs; \cite{2010IBVS.5927....1K}) discovered by
\cite{2003ApJ...583L..31G}. While sdBVr stars show short pulsation
periods ($\sim 80-400$ s) ascribed to radial modes and non-radial
$p$-modes, sdBVs pulsations (with periods $\sim 2500-7000$ s) are
associated to non-radial long period $g$-modes. Pulsations in both
groups of variable stars have been explained by the action of the
$\kappa$-mechanism due to the partial ionization of iron group
elements in the outer layers, where these elements are enhanced by the
action of radiative levitation \cite{1997ApJ...483L.123C}
\cite{2003ApJ...597..518F}.  While many sdBV stars are known, only one
He-sdB star, LS IV-14$^\circ$116, has been found to be a pulsator
\cite{2004Ap&SS.291..435A} \cite{2005A&A...437L..51A}. In fact, LS
IV-14$^\circ$116 is a very intriguing object: while its atmospheric
parameters ($T_{\rm eff}$ and $g$) place it well within the sdBVr
instability region
\cite{2010MNRAS.409..582N}\cite{2011ApJ...734...59G}, it displays
periods typical of the sdBVs family of pulsators
\cite{2005A&A...437L..51A} \cite{2011ApJ...734...59G}. From a
spectroscopic point of view, LS IV-14$^\circ$116 is also an intriguing
object, showing a mild He-enrichment ($n_{\rm He}=0.16$) and very high
excesses of zirconium, yttrium and strontium in its atmosphere
\cite{2011MNRAS.412..363N}. In this sense, both the He-enrichment and
the heavy metal excesses might be related to the effects of ongoing
diffusion before reaching diffusive equilibrium, thus placing the star
in the pre-EHB phase.  Up to now, the driving mechanism behind the
long period pulsations of LS IV-14$^\circ$116, as well as the absence
of short period pulsations, remains a mystery and the star poses a
challenge to the theory of stellar pulsations
\cite{2008ASPC..392..231F}.  In this connection, the recent
confirmation of both the multiperiodic variability and the $T_{\rm
  eff}-g$ values for LS IV-14$^\circ$116 \cite{2011ApJ...734...59G}
strongly increase the enigma. In a recent article
\cite{2011ApJ...741L...3M} we have suggested that pulsations in LS
IV-14$^\circ$116 might be excited by the $\epsilon$-mechanism acting
on unstable He-burning shells that take place before the begining of
the EHB.  In the $\epsilon$-mechanism, the driving is due to the
strong dependence of nuclear burning rates on temperature. In the
layers where nuclear reactions take place, thermal energy is gained at
compression by the enhancement of nuclear energy liberation, while the
opposite happens during the expansion phase
\cite{1989nos..book.....U}. In particular, in
\cite{2011ApJ...741L...3M} we have shown that the $\epsilon$-mechanism
is able to excite pulsations in hot-flasher stellar models. Although
only one hot-flasher sequence was analyzed in
\cite{2011ApJ...741L...3M}, it was shown that the $\epsilon$-mechanism
is qualitatively able to reproduce the long period ($P>1000$s)
$g$-modes at the values of $\log g$ and $\log T_{\rm eff}$ inferred
for LS IV-14$^\circ$116. As post-merger models also undergo
off-centered He-shell flashes before settling on the EHB
\cite{2000MNRAS.313..671S} and  given that  the location  of
the burning shell is particularly relevant for the range of excited
periods \cite{2011ApJ...741L...3M}, it seems natural to explore the
effect of the $\epsilon$-mechanism in post HeWD merger stellar models.

Interestingly enough, no star has been identified so far to be excited
by the epsilon mechanism. Aside from our previous
suggestion\cite{2011ApJ...741L...3M}, the only possible exceptions are
some oscillations in $\delta$ Scuti stars \cite{2006CoAst.147...93L},
PG 1159 stars \cite{1986ApJ...306L..41K} \cite{2009ApJ...701.1008C},
and more recently, some very long periods of Rigel
\cite{2012ApJ...749...74M}, but none of these have yet been 
confirmed.  In this connection, the identification of the
$\epsilon$-mechanism as the driving mechanism behind the lightcurve
variations of LS IV-14$^\circ$116 would offer the first evidence that the
$\epsilon$-mechanism can indeed drive pulsations in stars.

\section{Numerical methods and input physics}
\label{sec:NumMod}
%and \cite{RefJ}

Numerical models of 1D-post-merger structures were constructed by
simulating the merger process as an accretion process of the material
of the secondary on top of the primary (a He-WD model). Specifically
most of our models are constructed with {\tt MESA} stellar evolution
code \cite{2011ApJS..192....3P} following the recipe detailed in
\cite{2012MNRAS.419..452Z} for fast merger episodes in which the
accretion rate is of $10^4$M$_\odot$/yr. Also some variants of the
``composite merger'' (see \cite{2012MNRAS.419..452Z}) were
computed. It is worth noting that nucleosynthesis is included during
the accretion process.  The composition of the accreted material was,
by mass fraction, $^{12}$C=6$\times 10^{-5}$, $^{14}$N=1.27$\times
10^{-2}$, $^{16}$O=1.64$\times 10^{-3}$, $^{20}$Ne=1.88$\times
10^{-3}$ and $^{24}$Mg=3.68$\times 10^{-3}$ for the $Z=0.02$ sequences
and of $^{12}$C=3.1$\times 10^{-5}$, $^{14}$N=2.5$\times 10^{-3}$,
$^{16}$O=9.3$\times 10^{-5}$, $^{20}$Ne=3.9$\times 10^{-4}$ and
$^{24}$Mg=$8.0\times 10^{-4}$ for the $Z=0.004$ sequences.  Hydrogen
present in both the accretor and the accreted material was neglected.
Evolution during the helium shell sub-flashes that follow the
accretion process was followed with {\tt LPCODE} stellar evolution
code\cite{2005A&A...435..631A}, which has been extensively used in the
computation of He-flashes (e.g. \cite{2006A&A...449..313M} and
\cite{2008A&A...491..253M}). The pulsational instability analysis
presented in this work was carried out with the linear, nonradial,
nonadiabatic pulsation code described in detail in
\cite{2009ApJ...701.1008C} and references therein, in which the
$\epsilon$-mechanism for driving pulsations in the He-burning regions
is fully taken into account. As described in
\cite{2011ApJ...741L...3M} the ``frozen-in convection'' approximation
was asssumed because the timescales of the pulsations are usually much
shorter than the convective turnover times.

\subsection{Computed sequences}
\label{subsec:sequences}
\begin{table}
\caption{Global properties of the merger sequences studied in this
  work compared with those of LS IV-14$^\circ$116 and the Hot-Flasher
  sequence of \cite{2011ApJ...741L...3M}}
\label{tab:1}       % Give a unique label
\begin{tabular}{lllll}
\hline\noalign{\smallskip}
 & Stellar Mass & Initial Metallicity & Range of & H and He \\
  &(M$_\odot$)   &(Z$_0$) & excited periods & abundances by mass fraction\\
\noalign{\smallskip}\hline\noalign{\smallskip}
0.25M$_\odot$+0.25M$_\odot$ & 0.50M$_\odot$ & 0.02 & $<3700$s & without H \\
0.20M$_\odot$+0.20M$_\odot$ & 0.40M$_\odot$ & 0.02 & $<2600$s & without H \\
0.25M$_\odot$+0.15M$_\odot$ & 0.40M$_\odot$ & 0.02 & $<2700$s & without H \\
0.30M$_\odot$+0.10M$_\odot$ & 0.40M$_\odot$ & 0.02 & $<2500$s & without H \\
0.25M$_\odot$+0.19M$_\odot$ & 0.44M$_\odot$ & 0.004 & $ <2250$s & without H \\
0.35M$_\odot$+0.09M$_\odot$ & 0.44M$_\odot$ & 0.004 & $<3440$s & without H \\
0.25M$_\odot$+0.25M$_\odot$ & 0.50M$_\odot$ & 0.02 & $<3700$s & H=0.456 ,He=0.509 \\
0.25M$_\odot$+0.15M$_\odot$ & 0.40M$_\odot$ & 0.02 & $<2700$s & H=0.456 ,He=0.517 \\
\noalign{\smallskip}\hline\noalign{\smallskip}
Shallow Mixing \cite{2011ApJ...741L...3M} & 0.47378M$_\odot$ & 0.02 & $<2200$s & H=0.382 ,He=0.598 \\
{\bf  LS IV-14$^\circ$116}\cite{2011MNRAS.412..363N}  & - &  - &1954s-5084s & H=0.544 ,He=0.452\\
\noalign{\smallskip}\hline
\end{tabular}
\end{table}

\begin{figure}
%\resizebox{0.75\columnwidth}{!}{%
\resizebox{0.83\columnwidth}{!}{\includegraphics{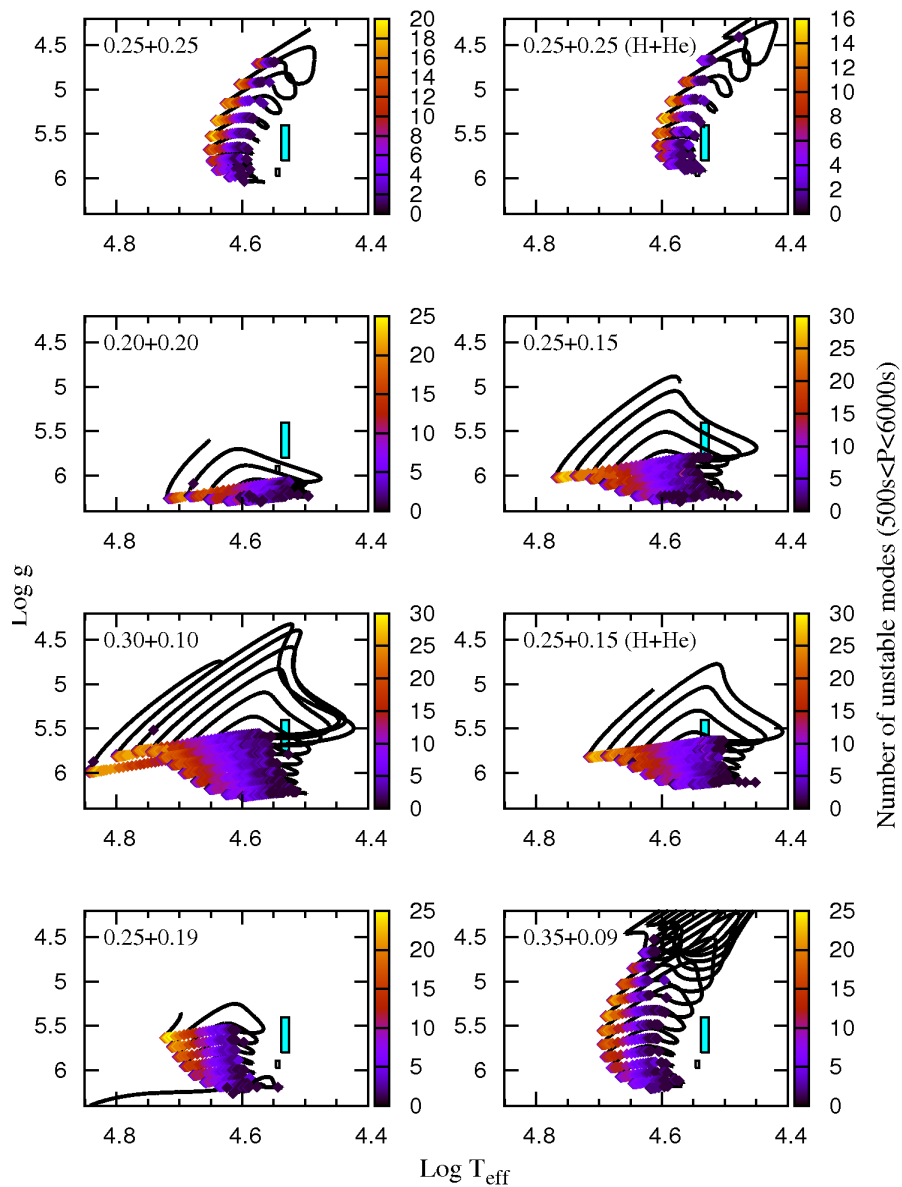} }
\caption{Location in the $\log g$-$\log T_{\rm eff}$ of the post-merger
  sequences studied in this work. Thick color points indicate the
  location of models for which pulsations driven by the
  $\epsilon$-mechanism have been found with periods in the studied
  range. Color coding indicates the number of unstable modes found.
  Note that in all the sequences unstable modes are found during the
  fast redward evolution during the He-shell flashes. Cyan and grey
  boxes indicate the derived atmosphere parameters of LS
  IV-14$^\circ$116 derived by \cite{2011MNRAS.412..363N} and
  \cite{2011ApJ...734...59G}, respectively.}
\label{fig:1}       % Give a unique label
\end{figure}

In order to study the excitation of $g$-modes through the
$\epsilon$-mechanism acting on the He-subflashes, and also to allow
for comparisons with LS IV-14$^\circ$116, we analysed the stability of
modes with periods between 500s and 6000s. Modes were analysed every 5
stellar evolution timesteps during the whole pre-EHB phase for all
evolutionary sequences.  The analysed sequences are detailed in Table
\ref{tab:1} and the location of the sequences in the $\log g$-$\log
T_{\rm eff}$ diagram is shown in Fig. \ref{fig:1}. In addition, we
analysed two sequences for which we added artificially a H-rich
envelope with a H/He mixture similar to that inferred in LS
IV-14$^\circ$116. In order to maximize the effect of the presence of
H, H was added as deep as possible without reaching layers with
temperatures high enough to burn H ($\log (1-m(r)/M_{\rm star})\sim
-3.85$). As will be shown, when comparing the $\log g$-$\log T_{\rm
  eff}$ of theoretical sequences and real stars, the total amount of H
in the envelope should be taken into account.

\begin{figure}
\resizebox{0.83\columnwidth}{!}{\includegraphics{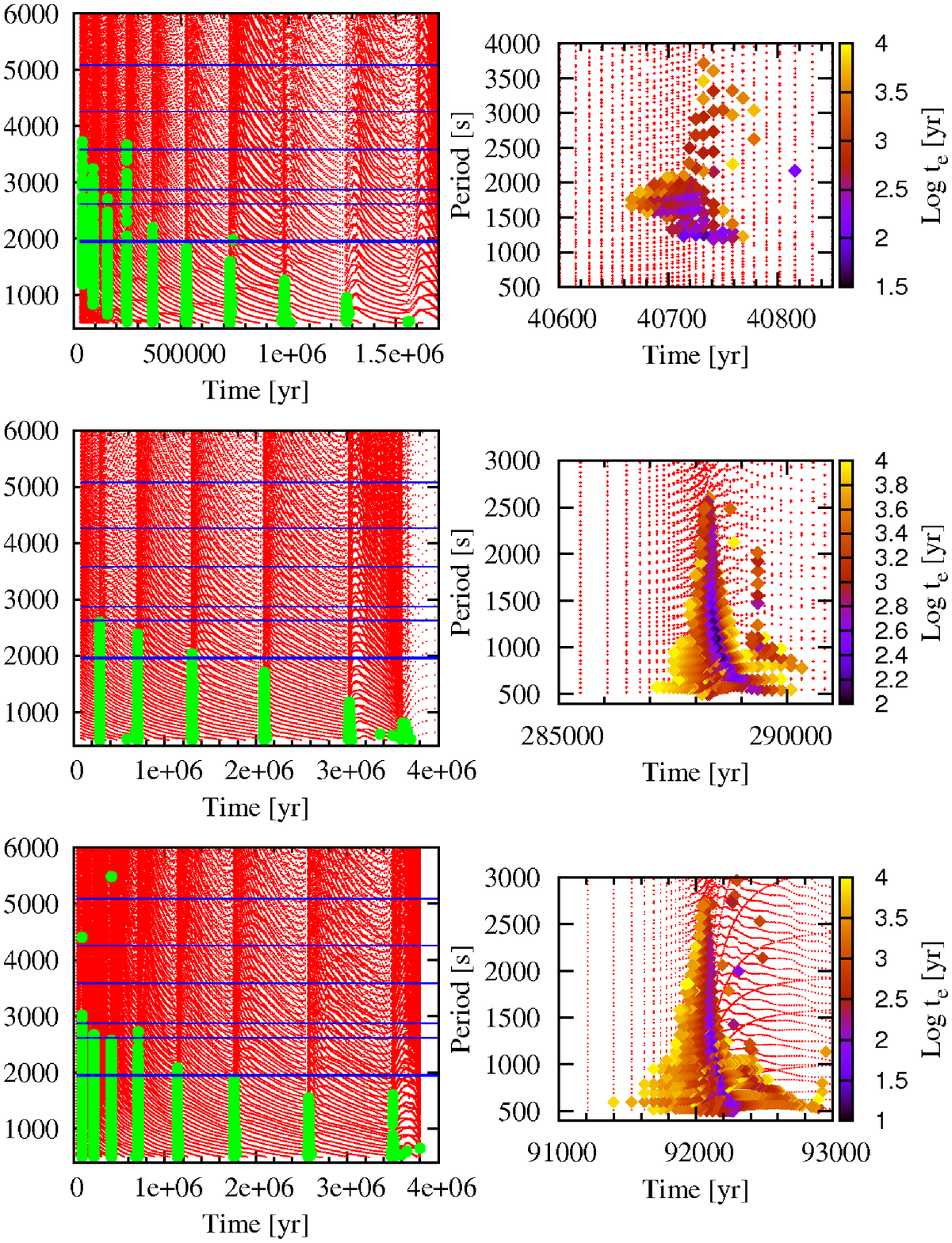} }
\caption{{\it Left:} Evolution of the periods of the normal modes
  during the pre-EHB evolution. Excited periods through the
  $\epsilon$-mechanism are marked with green thick points and compared
  with the observed modes of LS IV-14$^\circ$116 (blue horizontal
  lines). {\it Right:} Same as left Fig.s but only during the
  first (and most violent) flash. Color coding indicates the value of
  the e-folding time of the unstable periods. The sequences plotted are
  the $0.25$M$_\odot+0.25$M$_\odot$ merger ({\it top}), the
  $0.20$M$_\odot+0.20$M$_\odot$ merger ({\it middle}) and the
  $0.25$M$_\odot+0.15$M$_\odot$ merger with an added hydrogen
  envelope ({\it bottom}).}
\label{fig:2}       % Give a unique label
\end{figure}

\section{Discussion and Conclusions}
\label{sec:Conclu}
From Fig. \ref{fig:1} it is clear that more massive models evolve
during the He-core flashes at higher temperatures than less massive
ones. This is not unexpected as pure helium structures will settle on
the so-called He main sequence\cite{1994sse..book.....K} which lies at
higher temperatures for more massive models. Consequently the location
of our sequences of higher masses is unable to fit the observed
parameters of LS IV-14$^\circ$116.  It should be noted, however, that
when H is added in the envelope of the theoretical sequences, tracks
are shifted to lower temperatures and gravities, closer to the
observed location of LS IV-14$^\circ$116 (see top panels of
Fig. \ref{fig:1}). In the second and third row panels of
Fig. \ref{fig:1}, the evolution of the lower mass mergers analysed in
this work is shown ($M_{\rm final}=0.4$M$_\odot$). As can be seen,
lower mass models evolve in the pre-HB at lower temperatures (and
through wider loops) than the more massive ones attaining temperatures
similar to those of LS IV-14$^\circ$116. In addition, it can be seen
that the surface gravity of the sequences depend on how the final mass
is attained, with mergers with initially more massive primary stars
(i.e. lower accreted masses) evolving at lower gravities.

As shown in Table \ref{tab:1} and Fig. \ref{fig:2} the
$\epsilon$-mechanism in some of the more massive post-merger models is
able to excite periods as long as 3700s, significantly longer than
those predicted by the hot-flasher model studied in
\cite{2011ApJ...741L...3M}, making the predicted periods of the
unstable modes closer to those of LS IV-14$^\circ$116. It should be
noted however, that the formally unstable phase of the more massive
sequence during the first flash (top panel, Fig. \ref{fig:2}) is too
short, with the shortest e-folding times very similar to the length of
the instability phase ($\sim 100$yr). Then, although that sequence
displays formally unstable modes as long as 3700s they will not be
observable. The situation is different for the less massive sequences
(middle and bottom panels of Fig. \ref{fig:2}), for which the
instability phase is longer than the e-folding times, and thus
pulsations driven by the $\epsilon$-mechanism should be observed. It
is particularly worth noting that for the 0.25M$_\odot$+0.15M$_\odot$
sequence with H in the envelope (third row, right panel in
Fig. \ref{fig:1}), the remnant evolves through the observed location
of LS IV-14$^\circ$116 during the first three flashes, showing during
this stage excited periods as long as the two shortest periods
observed in LS IV-14$^\circ$116 ($P=1953.74,\,
2620.27$s\cite{2011ApJ...734...59G}). In particular, the range of
excited periods in the first five flashes is able to reproduce the
higher amplitude mode of LS IV-14$^\circ$116 ($P=1953.74$). Still, the
excited modes of all our computed sequences have periods too short to
explain the longest periods observed in LS IV-14$^\circ$116
($P>4000$s). It should be noted, however, that the merger models we
have adopted in this study are based on some simplifying assumptions,
and thus more detailed merger models should be explored before making
a final statement about the origin of LS IV-14$^\circ$116 and the
mechanism behind its pulsations. In particular, it is worth noting
that the thermal structure of post-merger models is sensitive to the
details of the merger process and on the relative masses of the
primary and secondary stars\cite{2009A&A...500.1193L}.  A more
realistic treatment of the merger process might lead to thermal
structures signifficantly different and thus to a shift in the
location of the He-core flashes, altering the range of excited
periods.  One interesting feature predicted by the present explanation
of the pulsations observed in LS IV-14$^\circ$116 is the large
magnitude of the rate of period changes of the modes. In fact, as the
structure changes very fast during the helium core flashes, then the
predicted period change can be as high as a few seconds per year. The
determination of large value in the rate of period changes would be a
strong indication that LS IV-14$^\circ$116 is in a fast evolutionary
stage. In this connection it might be worth noting that the value of
the determined period of the $\sim 2870$s mode of LS IV-14$^\circ$116
differs in about 2s from the observations done in 2004
\cite{2005A&A...437L..51A} and the determinations of 2010
\cite{2011ApJ...734...59G}, which is somewhat larger than the quoted
error bars ($\sigma<1$s).

Despite the particular origin of LS IV-14$^\circ$116, and it is by no
means obvious whether it is a post-merger object or not, our analysis
suggest that post HeWD+HeWD merger remnants should go through a phase
in which pulations are driven by the $\epsilon$-mechanism. Due to the
short duration of the unstable phases, only $\sim 1$ every 100 pre-EHB
objects should be pulsators.  A more detailed exploration of the
parameter space of both the merger and hot-flasher scenarios to
understand the pulsations excited in LS IV-14$^\circ$116 will be
performed in forthcoming works. We expect that future work will help
to distinguish between the possible evolutionary channels for LS
IV-14$^\circ$116.

{\it Acknowledgments:} M3B thanks the organizers of the ``40th Liege
International Astrophysical Colloquium'' for the finantial assistance
that helped him to attend the conference.  This research was supported
by  PIP 112-200801-00940 from CONICET and
PICT-2010-0861 from ANCyT.

%\begin{thebibliography}{}

\bibliographystyle{epj}

\bibliography{mmiller}

%Author, Journal \textbf{Volume}, (year) page numbers
%{2000MNRAS.313..671S}
%{2001ApJ...562..368B}
%{2010IBVS.5927....1K}

%Author, \textit{Book title} (Publisher, place year) page numbers

% etc
%\end{thebibliography}

\end{document}